\documentclass[aps,prb,twocolumn,groupedaddress,showpacs]{revtex4}
\usepackage{epsfig}
\def\be{\begin{equation}}
\def\ee{\end{equation}}
\def\bea{\begin{eqnarray}}
\def\eea{\end{eqnarray}}
\def\bc{\begin{center}}
\def\ec{\end{center}}
\usepackage{longtable}
\usepackage{psfig}

\begin{document}

\title{Nature of quasiparticle excitations in the fractional quantum Hall effect}
\author{Gun Sang Jeon and Jainendra K. Jain}
\affiliation{Physics Department, 104 Davey Laboratory, The Pennsylvania State University,
University Park, Pennsylvania 16802}
\date{\today}

\begin{abstract}
We investigate the 1/3 fractional quantum Hall state with one and two 
quasiparticle excitations.
It is shown that the quasiparticle excitations are best described as 
excited composite fermions occupying higher composite-fermion quasi-Landau levels.
In particular, the composite-fermion wave function for a single 
quasiparticle has 15\% lower energy than the trial wave function  
suggested by Laughlin, and for two quasiparticles, the 
composite fermion theory also gives new qualitative structures.
\end{abstract}

\maketitle

\section{Introduction}

Interacting electrons in two dimensions form a collective  
quantum fluid~\cite{FQHE} under the application of an intense
magnetic field that exhibits the remarkable 
phenomenon of the fractional quantum Hall 
effect(FQHE).~\cite{FQHEexp}
An early trial wave function proposed by Laughlin~\cite{Laughlin}
for the ground state at the filling factor $\nu=1/m$, $m$ odd,
turned out to work well.  The composite-fermion (CF) 
theory~\cite{CF} applies to a broader range of phenomena, while also
providing a new interpretation for the physics of the $\nu=1/m$ state, 
as a state of composite fermions at an effective filling of $\nu^*=1$.  
While the wave function for the $\nu=1/m$ ground state from the CF 
theory is the same as that in Ref.~\onlinecite{Laughlin}, the wave 
functions for the excitations are different, 
which gives an opportunity to test the validity of the CF theory
at $\nu=1/m$ itself.

We note that when speaking of ``quasiparticles" in this paper, 
we really will mean ``quasiparticle excitations" of an incompressible 
FQHE state.  In a broader sense, the quasiparticles, namely the objects 
that are weakly 
interacting,  are known to be composite fermions; many  
properties of the FQHE state and various related phenomena can be 
understood in terms of almost free composite fermions.~\cite{FQHE}  
We will see that the quasiparticle excitations are also nothing but  
the familiar composite fermions.

Different candidate theories for the 
charged quasiparticle excitations of the incompressible quantum Hall states
have been studied in the past.~\cite{Laughlin,CF,Morf,Kasner,Girlich,Bonesteel} 
Two competing views are as follows.  In Ref.~\onlinecite{Laughlin} the quasiparticle 
was thought of as a ``vortex," based on which a wave function was proposed.
In Ref.~\onlinecite{CF}, the quasiparticle was viewed as an excited composite
fermion, which produces a different wave function.
Many studies~\cite{Kasner,Girlich,Bonesteel} have compared the two 
wave functions, and, irrespective of the geometry and the type of interaction
employed in these studies,
the CF wave function has been found to be superior.

Our objective in this work is 
to compare the two theories for systems containing more than one quasiparticle.
A {\em qualitative} difference between the two approaches has been 
noted in recent microscopic studies of multi-quasiparticle 
states~\cite{Kjonsberg,fracstat} in the context of fractional statistics;
the statistics parameter, which is a statement 
regarding the winding properties of quasiparticles around one another, 
was found to possess a definite value only for the CF wave functions.
Our present study finds qualitative differences as well as substantial 
quantitative deviations between the two approaches.

\section{Single quasiparticle}

We will concentrate below on $\nu=1/3$, and 
begin, for completeness, with one quasiparticle in the disc geometry.
We will assume 
Coulomb interaction and investigate systems with up to 160 electrons.

The trick of piercing a flux quantum
adiabatically through the system motivates the following wave functions for the 
quasiparticle~\cite{Laughlin} 
\be
\Psi_L^{1qp} = 
e^{ -\sum_j |z_j|^2/4} 
\prod_l \left( 2 \frac{\partial}{\partial z_l} \right) 
\prod_{j<k} (z_j - z_k)^3 , 
\label{L1qp}
\ee
where the position of the $j$th electron is denoted by $z_j\equiv x_j - i
y_j$, and the length is measured in units of the magnetic length $\ell \equiv
\sqrt{\hbar c/eB}$.
This wave function describes a quasiparticle located at the origin.

\begin{figure}
\centerline{\epsfig{file=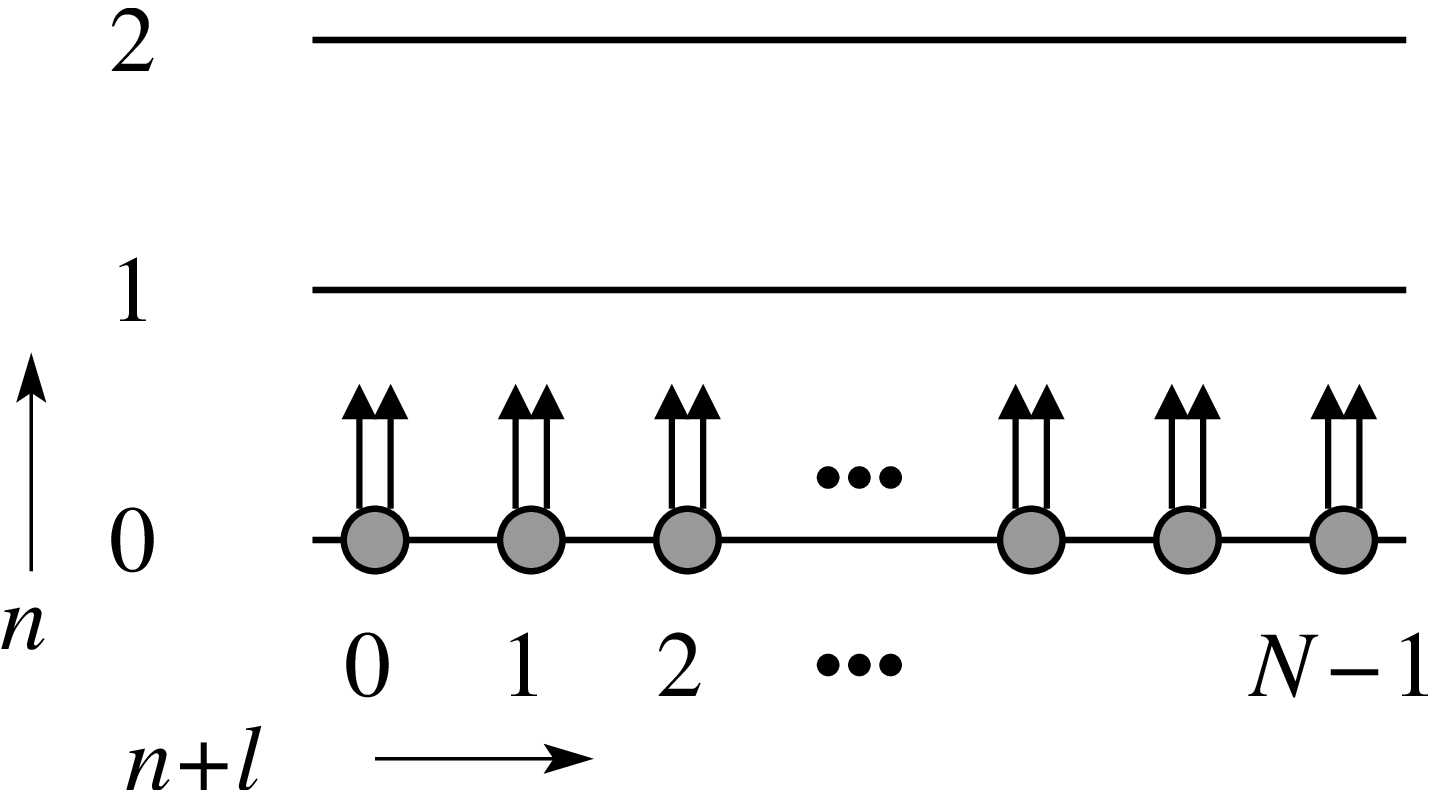,width=3.0in,angle=0}}
\centerline{(a)}
\vspace*{5mm}
\centerline{\epsfig{file=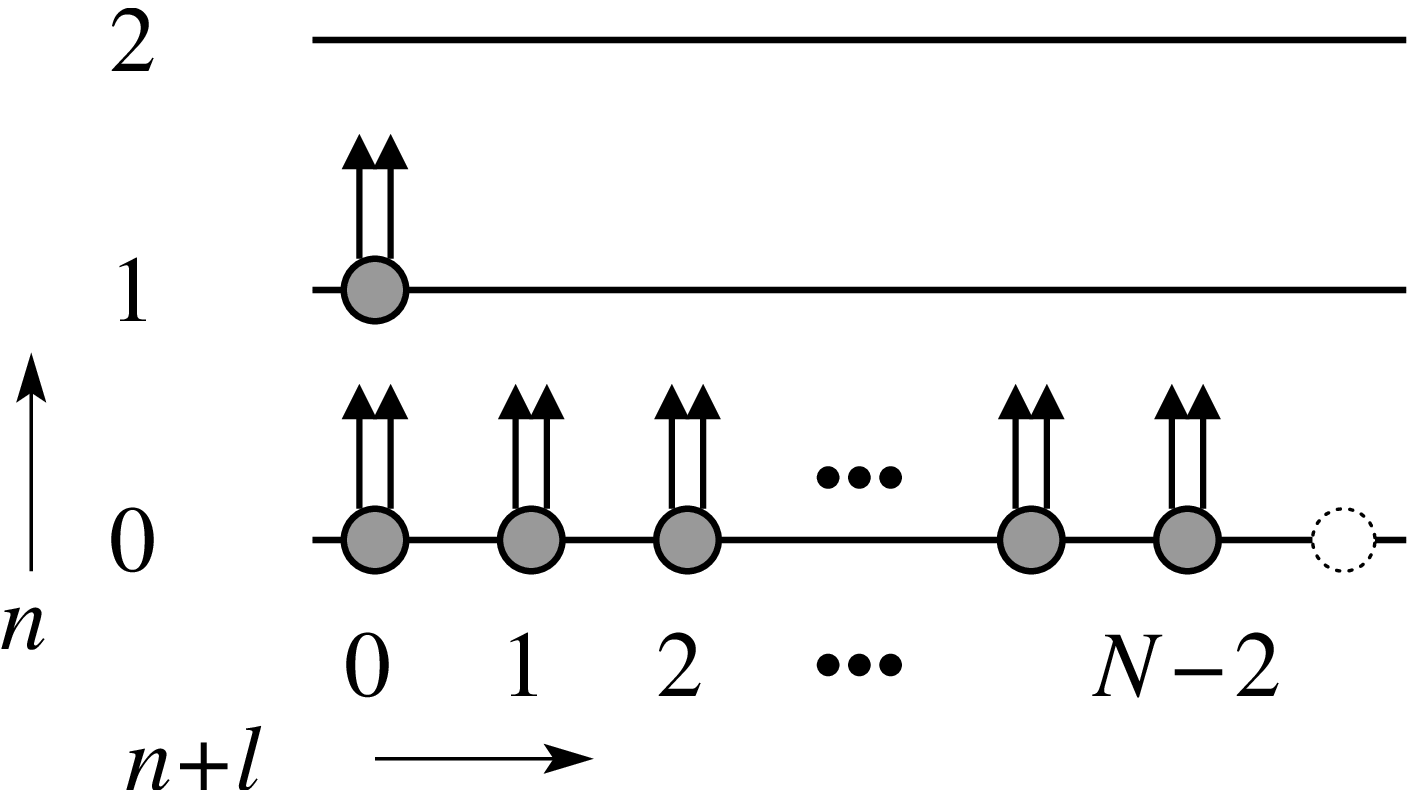,width=3.0in,angle=0}}
\centerline{(b)}
\caption{
	Schematic depiction of (a) the ground state, and 
	(b) the single quasiparticle excitation at $\nu=1/3$ 
	in terms of composite fermions.
	The composite fermions are shown as electrons with two arrows, 
	where the arrows represent the vortices bound to composite
	fermions.  The labels 
	$n$ and $l$ denote the CF quasi-Landau level index and the angular
	momentum of the composite fermion, respectively.
}
\label{fig:CF1QP}
\end{figure}

In the CF theory, the wave functions for the ground as well as excited
states are constructed by analogy to the corresponding wave functions at 
an effective filling factor, which is 
$\nu^*=1$ in the case of $\nu=1/3$.   
These are shown schematically in Fig.~\ref{fig:CF1QP}.
The ground state for $\nu=1/3$ takes the form~\cite{CF} 
\be
\Psi^{GS} = {\cal P} \prod_{j<k} (z_j - z_k)^2 \Phi_1,
\ee
where $\Phi_1$ represents the incompressible integral
quantum Hall ground state at filling factor unity, and 
${\cal P}$ is an operator that projects the state onto the
lowest Landau level(LL).
This wave function is identical to the ground state trial wave
function in Ref.~\onlinecite{Laughlin}:
\be
\prod_{j<k} (z_j - z_k)^3 e^{ -\sum_i |z_i|^2/4}\;.
\ee
Further extending the analogy, 
the quasiparticle is analogous to the integral quantum Hall 
state in which the lowest LL is completely occupied and the 
second has precisely one electron. [See Fig.~\ref{fig:CF1QP}(b).]
The CF wave function for the quasiparticle at $\nu=1/3$ is then given by 
\bea
\Psi_{CF}^{1qp} &=& {\cal P} \prod_{j<k} (z_j - z_k)^2 
\left|
\begin{array}{ccc}
z_1^* & z_2^* & \ldots\\
1 & 1 & \ldots\\
z_1 & z_2 & \ldots\\
\vdots & \vdots&\ldots\\
z_1^{N-2} & z_2^{N-2} & \ldots
\end{array}
\right|
\nonumber \\
&& \times \exp \left[ -\frac14 \sum_j |z_j|^2 \right]. 
\eea
The lowest LL projection is accomplished by bringing all of the $z^*_j$'s to the left,
and replacing $z^*_j\rightarrow 2\partial/\partial z_j$, with the understanding
that the derivatives do not act on the Gaussian factor.~\cite{JK} This produces 
\begin{eqnarray}
\Psi_{CF}^{1qp}&=&
e^{-\sum_{i}|z_{i}|^{2}/4} 
\left[\sum_{i=1}^{N} \frac{2{\partial\over \partial z_i}}{\prod_j^{'}
(z_j-z_i)}\right] \prod_{l<m}(z_l-z_m)^3 \nonumber \\
&=&  \sum_{i=1}^{N} \frac{6\sum_k^{'}(z_i-z_k)^{-1}}
{\prod_j^{'}(z_j-z_i)} \Psi^{GS},
\end{eqnarray}
where the prime denotes the condition $j\neq i$, or $k\neq i$. 
$\Psi_{CF}^{1qp}$ is different from $\Psi_{L}^{1qp}$.
The lowest LL projected form of this wave function is fairly
complicated and would have been difficult to guess without guidance from the CF theory.
The presence of factors of the type $(z_j-z_i)$ in the
denominator is not a problem because they are canceled by similar factors in the
numerator.

In order to investigate which wave function gives a better description
of the quasiparticle, we compute the expectation values of 
the Coulomb energy for each state: 
\bea
E_L^{1qp} &\equiv& 
\frac{\langle \Psi_L^{1qp} | V | \Psi_L^{1qp}\rangle}
{\langle \Psi_L^{1qp} | \Psi_L^{1qp}\rangle}, \label{EL}\\
E_{CF}^{1qp} &\equiv& 
\frac{\langle \Psi_{CF}^{1qp} | V | \Psi_{CF}^{1qp}\rangle}
{\langle \Psi_{CF}^{1qp} |  \Psi_{CF}^{1qp}\rangle, \label{ECF}}
\eea
with the total Coulomb energy 
\begin{equation}
V \equiv \sum_{j<k} \frac{1}{|z_j - z_k|}\;.
\end{equation}
(Energies will be measured in units of $e^2/\ell$.)
Since both trial wave functions are in the lowest LL, 
and the ground state is the same,
the difference 
\begin{equation}
\Delta E^{1qp} \equiv E_L^{1qp} - E_{CF}^{1qp}
\end{equation}
gives the energy difference between the quasiparticle energies 
predicted by the two theories.

$\Psi_L^{1qp}$ has many derivatives with
respect to the coordinates, the number of which increases with the
number of electrons; this does not allow a direct evaluation of
$E_L^{1qp}$ via Monte Carlo sampling according to $|\Psi_L^{1qp}|^2$. 
However, an evaluation of $E_L^{1qp}$ in Eq.~(\ref{EL}) is 
possible by a method similar to that used in Ref.~\onlinecite{Morf}.
The partial integration of all derivatives reduces
the denominator and the numerator, respectively, to the forms 
\bea
\langle \Psi_L^{1qp} | \Psi_L^{1qp}\rangle
= \int \prod_k dz_k dz_k^* |\Psi^{GS}|^2 \prod_k (|z_k|^2-2), \\
\langle \Psi_L^{1qp} |V| \Psi_L^{1qp}\rangle
= \int \prod_k dz_k dz_k^* \widetilde{V}
|\Psi^{GS}|^2 \prod_k (|z_k|^2-2)
\eea
with 
\bea \nonumber
\widetilde{V} &\equiv& \sum_{j<k} 
\Big[\frac{1}{|z_j-z_k|} + \frac{1}{|z_j-z_k|^5 (|z_j|^2-2)(|z_k|^2-2)}\\ 
\nonumber
&& \quad \times \Big\{ 9 + 2 |z_j-z_k|^2-3|z_j-z_k|^4 \\
&&  \quad \quad - 2 (|z_j-z_k|^2+3) {\rm Re}[z_j z_k (z_j^*-z_k^*)
	]\Big\}\Big].
\eea
Defining $S \equiv {\rm sgn} [\prod_k (|z_k|^2 -2 )]$, we can evaluate
$E_L^{1qp}$ by the expression
\be \label{eq:EL1QP}
E_L^{1qp} = \frac{\langle\!\langle S\widetilde{V} \rangle\!\rangle}
{\langle\!\langle S\rangle\!\rangle},
\ee
where $\langle\!\langle \ldots \rangle\!\rangle$ denotes the Monte
Carlo average with the weighting function 
$|\Psi^{GS}|^2 \prod_k \left| |z_k|^2 -2 \right|$. 
As a test, for $N=3$, we obtained via Eq.~(\ref{eq:EL1QP})
	$E_L^{1qp} = 0.89121 \pm 0.00002$, 
which is in good agreement with the exact value
$E_L=(12357/24576)\sqrt{\pi}\approx 0.8912033$.

\begin{figure}
\centerline{\epsfig{file=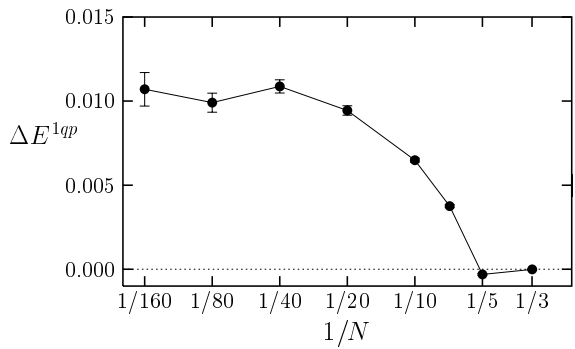,width=3.0in,angle=0}}
\caption{
Energy difference, in units of $e^2/\ell$, between $\Psi_L^{1qp}$ and 
$\Psi_{CF}^{1qp}$ for a single quasiparticle at the origin.}
\label{fig:E1QP}
\end{figure}

Figure~\ref{fig:E1QP} shows the energy difference $\Delta E^{1qp}$ as a
function of the electron number $N$ up to $N=160$.
For small numbers of electrons $(N \le 5)$, the energy difference
is nearly zero.
As $N$ is increased further, the deviation between the two energies 
begins to grow.  $\Delta E^{1qp}$ increases gradually, eventually saturating 
at a finite value $\approx 0.011$.
This result is consistent with previous studies, which
also reached the conclusion that the CF wave function has
lower energy.~\cite{Kasner,Girlich,Bonesteel}
It is of interest to note that the energy of $\Psi_L^{1qp}$  
is higher than that of $\Psi_{CF}^{1qp}$ by approximately 15\% of the
quasiparticle energy ($\sim 0.07$).  
(We note that an earlier calculation~\cite{Bonesteel} on the sphere obtained 
a difference of $\approx 0.005$ between the two quasiparticle 
wave functions; the reason for the discrepancy is unclear.
We cannot compare our results directly with 
those in Ref.~\onlinecite{Kasner} as that work employed
a model with short range interaction potential.)

\section{two quasiparticles}

\begin{figure}
\centerline{\epsfig{file=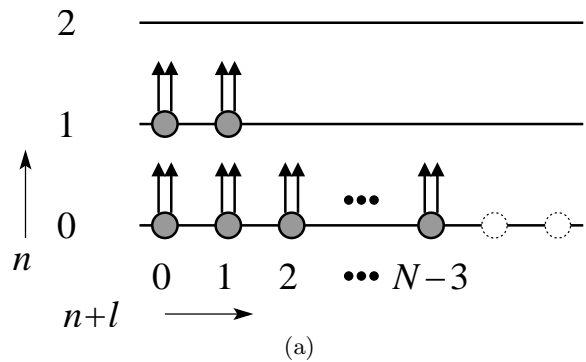,width=3.0in,angle=0}}
\centerline{(a)}
\vspace*{5mm}
\centerline{\epsfig{file=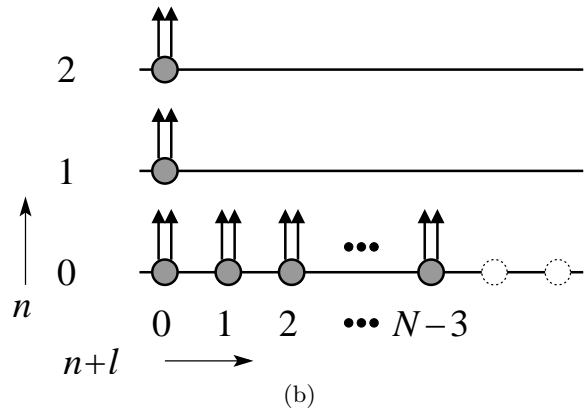,width=3.0in,angle=0}}
\centerline{(b)}
\caption{
	Schematic depiction of composite-fermion states (a) $[N-2,2]$ and  
	(b) $[N-2,1,1]$.  The former has two composite fermions in the second
	CF quasi-Landau level, whereas the latter has one in the second and one in the third.
	The labels $n$ and $l$ denote the CF quasi-Landau level index and the angular
        momentum of the composite fermion, respectively.
}
\label{fig:CF2QP}
\end{figure}

Next we consider states with two quasiparticles.
The CF theory provides the following wave function for two
quasiparticles at the origin~\cite{CF,JK,Dev} 
\bea
\Psi_{CF}^{[N-2,2]} &=& {\cal P} \prod_{j<k} (z_j - z_k)^2 
\left|
\begin{array}{ccc}
z_1^* & z_2^* & \ldots\\
z_1^* z_1 & z_2^* z_2 & \ldots\\
1 & 1 & \ldots\\
z_1 & z_2 & \ldots\\
\vdots & \vdots&\ldots\\
z_1^{N-3} & z_2^{N-3} & \ldots
\end{array}
\right|
\nonumber \\
&& \times \exp \left[ -\frac14 \sum_j |z_j|^2 \right]. 
\eea
Here the superscript $[N{-}2,2]$ represents the occupation number in
each CF quasi-Landau level, namely, $N{-}2$ electrons for $n=0$ and $2$ electrons for
$n=1$ CF quasi-Landau levels, as shown schematically in Fig.~\ref{fig:CF2QP}(a).
We will not show the explicit lowest LL projected form here, which is 
significantly more complicated than that for a single CF quasiparticle.

A generalization of Eq.~(\ref{L1qp}) to two quasiparticles 
is also straightforward:
\be
\Psi_L^{2qp} = 
e^{-\sum_j |z_j|^2/4} 
\prod_l \left( 2 \frac{\partial}{\partial z_l} \right) 
\left( 2 \frac{\partial}{\partial z_l} \right) 	
\prod_{j<k} (z_j - z_k)^3 
\ee
which is identical to the wave function considered in 
Ref.~\onlinecite{Kjonsberg} for two quasiparticles at the origin.

As for one quasiparticle, we have computed the difference
\be
\Delta E^{2qp}\equiv E_L^{2qp}-E_{CF}^{[N-2,2]} ,
\ee 
where
\bea
E_L^{2qp} &\equiv& 
\frac{\langle \Psi_L^{2qp} | V | \Psi_L^{2qp}\rangle}
{\langle \Psi_L^{2qp} | \Psi_L^{2qp}\rangle}, \\
E_{CF}^{[N-2,2]} &\equiv& 
\frac{\langle \Psi_{CF}^{[N-2,2]} | V | \Psi_{CF}^{[N-2,2]}\rangle}
{\langle \Psi_{CF}^{[N-2,2]} |  \Psi_{CF}^{[N-2,2]}\rangle }.
\eea

For composite fermions, the energy can be calculated 
using the methods described previously~\cite{JK} for 
fairly large systems.
The evaluation of the Coulomb energy for $\Psi_L^{2qp}$
involves similar difficulty as the energy of a single 
quasiparticle, caused by the presence of many derivatives.
What makes matters worse here is that
$\Psi_L^{2qp}$ has twice as many derivatives as $\Psi_L^{1qp}$,
producing more singular terms in $\widetilde{V}$.
Consequently, the convergence of the integration becomes poor and
the method employed in the preceding section turns out not
to be a clever way to treat $\Psi_L^{2qp}$.
As an alternative method, we have expanded $\Psi_L^{2qp}$ in the series
\bea 
\Psi_L^{2qp} &=& \sum_{\{l_j\},\{m_j\},\{n_j\}}'
C(\{l_j\},\{m_j\},\{n_j\}) D(l_1,\ldots,l_N)  
	\nonumber \\
&& \times D(m_1,\ldots,m_N) D(n_1,\ldots,n_N)
	\nonumber \\ \label{eq:expansion}
&& \times \exp \left[ -\frac14 \sum_j |z_j|^2 \right],
\eea
where 
\be \label{eq:det}
D(n_1,\ldots,n_N) \equiv 
\left|
\begin{array}{ccc}
\left(\frac{\partial}{\partial z_1}\right) ^{n_1} 1 & 
\left(\frac{\partial}{\partial z_2}\right)^{n_2} 1 & 
\ldots \\
\left(\frac{\partial}{\partial z_1}\right)^{n_1} z_1 & 
\left(\frac{\partial}{\partial z_2}\right)^{n_2} z_2 & 
\ldots \\
\vdots & \vdots & \ldots \\
\left(\frac{\partial}{\partial z_1}\right)^{n_1} z_1^{N-1} & 
\left(\frac{\partial}{\partial z_2}\right)^{n_2} z_2^{N-1} & 
\ldots 
\end{array}
\right|
\ee
and the primed summation runs under the restriction: $l_j, m_j, n_j \ge 0$
and $l_j+m_j+n_j = 2$ for all $j$.
The expansion in Eq.~(\ref{eq:expansion}) is possible because the
polynomial part in $\Psi^{GS}$ is the cube of a determinant.
It has an advantage over the previous method; we can obtain the analytic form 
of $\Psi_L^{2qp}$ without the explicit derivatives, which allows a  
direct evaluation of the wave function itself.
It is also seen that the determinant in Eq.~(\ref{eq:det}) vanishes
when all of $n_i$'s are nonzero.
For $N=3$ the numerical calculation using the wave function in
Eq.~(\ref{eq:expansion}) produced $E_L^{2qp}=1.2047\pm0.0001$,
which compares well with the exact 
value $(87/128)\sqrt{\pi} \approx 1.204715$.

\begin{figure}
\centerline{\epsfig{file=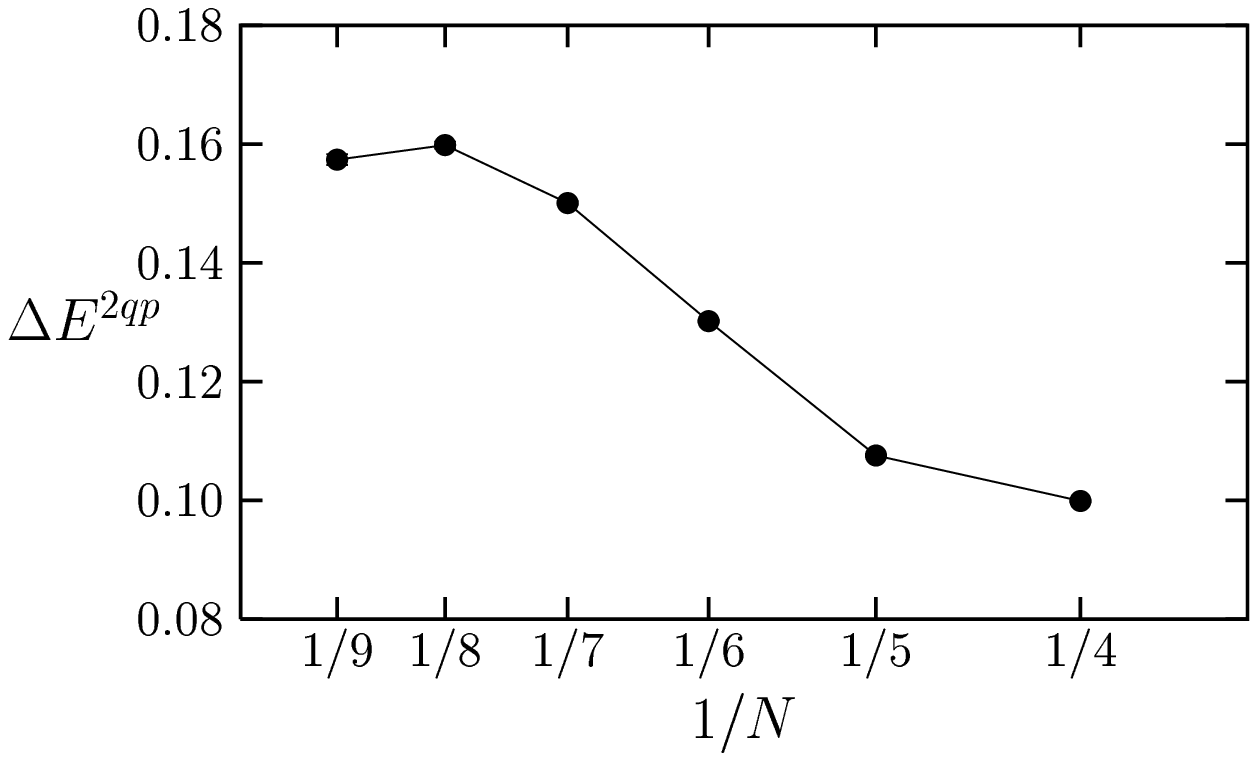,width=3.0in,angle=0}}
\centerline{(a)}
\vspace*{5mm}
\centerline{\epsfig{file=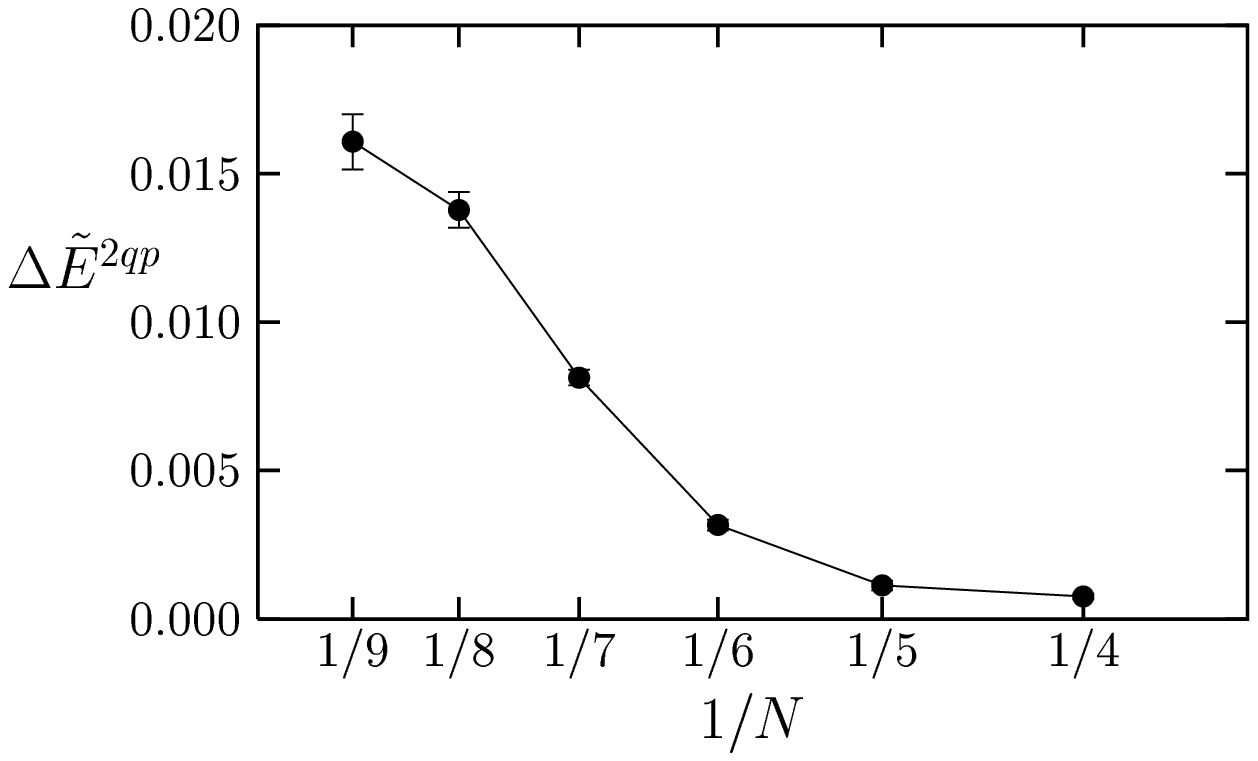,width=3.0in,angle=0}}
\centerline{(b)}
\caption{
Energy difference between (a) $\Psi_L^{2qp}$ and 
$\Psi_{CF}^{[N{-}2,2]}$, and (b) $\Psi_L^{2qp}$
and $\Psi_{CF}^{[N{-}2,1,1]}$. 
}
\label{fig:E2QP}
\end{figure}
Since the number of non-vanishing terms in Eq.~(\ref{eq:expansion})
increases rapidly, the integer coefficients 
$C(\{l_j\}, \{m_j\}, \{n_j\})$ were obtained numerically up to $N=9$.
We plot $\Delta E^{2qp}$ as a function of $N$ in Fig.~\ref{fig:E2QP}(a).
It is observed that $\Psi_L^{2qp}$ has a much higher energy
than $\Psi_{CF}^{[N-2,2]}$ even for a small number of electrons, unlike 
for a single quasiparticle.
This discrepancy increases as the system grows.
>From the closeness of the values for $N=8$ and $9$ we presume
that the thermodynamic limit for the energy 
difference is $\approx 0.16$, which is surprisingly 
big, even larger than the gap to creating a quasiparticle quasihole pair 
out of the ground state (which is roughly 0.1).
It far exceeds the naive estimate $2\Delta E^{1qp} \approx 0.022$.

\begin{figure}
\centerline{\epsfig{file=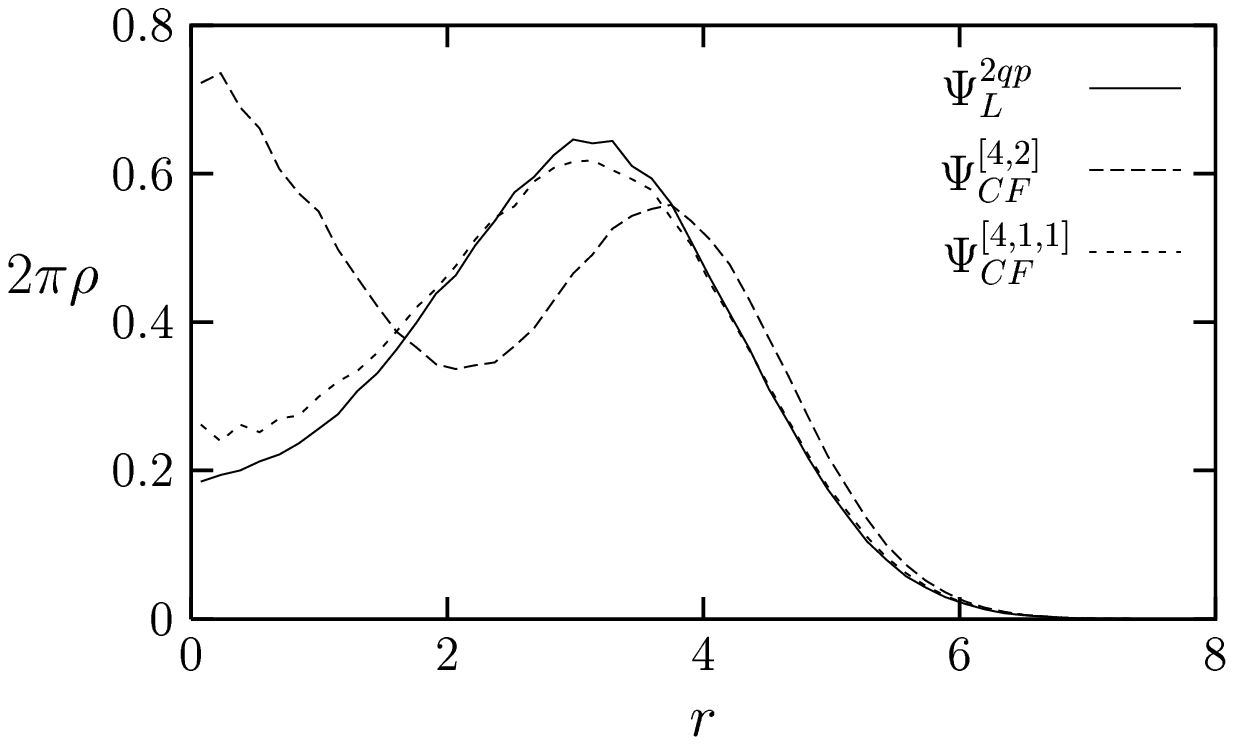,width=3.0in,angle=0}}
\centerline{(a)}
\vspace*{5mm}
\centerline{\epsfig{file=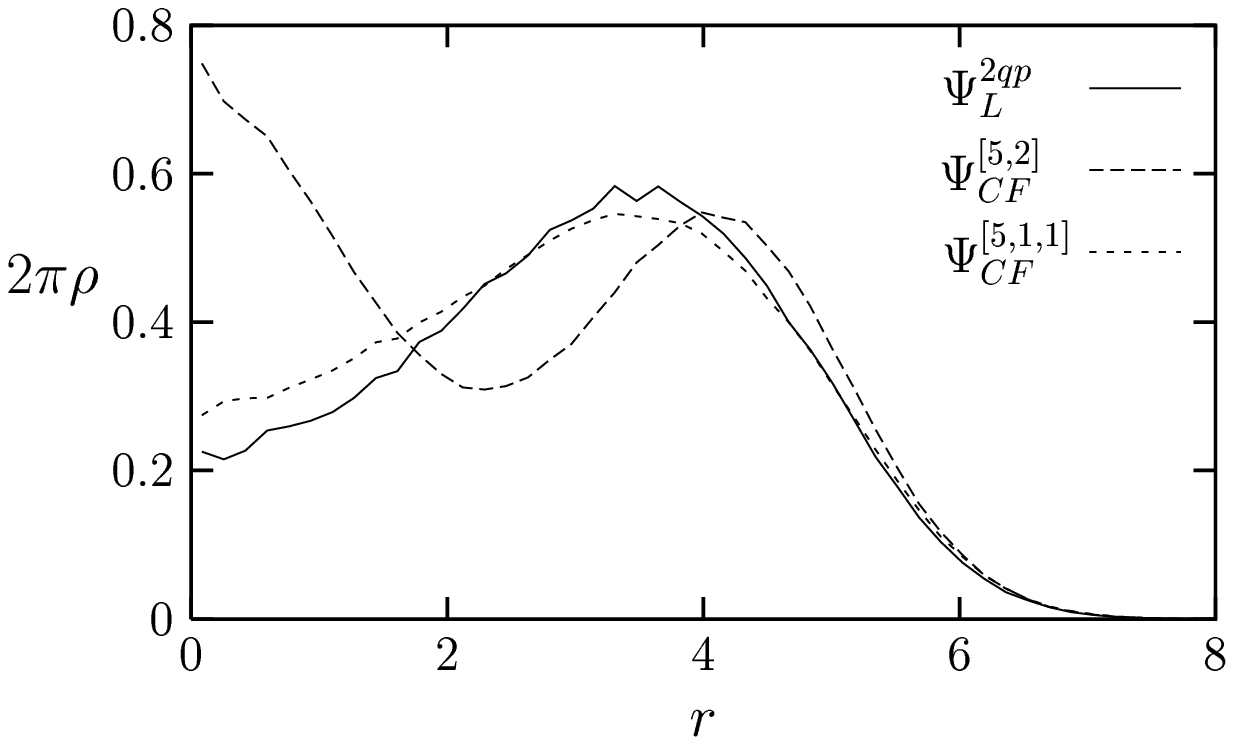,width=3.0in,angle=0}}
\centerline{(b)}
\caption{
Density profiles of $\Psi_L^{2qp}$,
$\Psi_{CF}^{[N{-}2,2]}$, and $\Psi_{CF}^{[N{-}2,1,1]}$
for (a) $N=6$, and (b) $N=7$.
}
\label{fig:rho2QP}
\end{figure}

A qualitative difference between the two approaches appears 
at the level of two quasiparticles.
The two wave functions considered above 
have different total angular momenta: 
\bea
M_{CF}^{[N-2,2]}&=& \frac12 (3N^2-7N+4),\\
M_L^{2qp}&=& \frac12 (3N^2-7N).
\eea
(The largest occupied single electron orbital has the same 
angular momentum for the two states, though.) 
There is no obvious way of constructing a generalization of 
$\Psi_L^{1qp}$ with an angular momentum $M_{CF}^{[N-2,2]}$.  
However, it turns out that the CF theory contains 
an {\em excited} state for two quasiparticles with  
angular momentum $M_L^{2qp}$:
\bea
\Psi_{CF}^{[N-2,1,1]} &=& {\cal P} \prod_{j<k} (z_j - z_k)^2 
\left|
\begin{array}{ccc}
(z_1^*)^2 & (z_2^*)^2 & \ldots\\
z_1^* & z_2^* & \ldots\\
1 & 1 & \ldots\\
z_1 & z_2 & \ldots\\
\vdots & \vdots&\ldots\\
z_1^{N-3} & z_2^{N-3} & \ldots
\end{array}
\right|
\nonumber \\
&& \times \exp \left[ -\frac14 \sum_j |z_j|^2 \right] \;.
\eea
This wave function contains one excited composite fermion in the second CF
quasi-Landau level but the other in the third CF quasi-Landau level, as 
indicated by the notation $[N{-}2,1,1]$.  See Fig.~\ref{fig:CF2QP}(b).
The energy of $\Psi_{CF}^{[N-2,1,1]}$ is expected to be 
higher than that of $\Psi_{CF}^{[N-2,2]}$, since the former has larger 
``kinetic energy'' for the composite fermions, which is of the order 
of the 1/3 gap.
We have also computed $\Delta \tilde{E}^{2qp} = E_L^{2qp} -
E_{CF}^{[N-2,1,1]}$ with $E_{CF}^{[N-2,1,1]} \equiv 
\langle \Psi_{CF}^{[N-2,1,1]} | V | \Psi_{CF}^{[N-2,1,1]}\rangle/
\langle \Psi_{CF}^{[N-2,1,1]} |  \Psi_{CF}^{[N-2,1,1]}\rangle $
as a function of the system size.
As shown in Fig.~\ref{fig:E2QP}(b), the energy difference is 
reduced compared with $\Delta E^{2qp}$.
This suggests that $\Psi_L^{2qp}$ actually resembles the CF
$[N{-}2,1,1]$ state.
That is also verified in the density plot of three states, which is
shown in Fig.~\ref{fig:rho2QP}.

It is obvious that as more and more quasiparticles are created by
repeated application of 
$ \prod_l ( 2 \partial/\partial z_l )$,
the resulting state will contain, in the CF interpretation, composite
fermions in higher and higher CF quasi-Landau levels; this state will have much higher
energy than the state with all CF quasiparticles in the second 
CF quasi-Landau level.

\section{Conclusion}

Our study confirms the description of quasiparticles as 
composite fermions in an excited composite-fermion quasi-Landau level.
This physics not only gives the best available microscopic wave 
functions for the quasiparticles, but also brings out 
new qualitative structures for multi-quasiparticle states.
While we have concentrated on the vicinity of $\nu=1/3$ 
above, it should be noted that the CF theory explains 
the general phenomenon of the FQHE, giving accurate wave functions
for the ground states,  quasiparticles, quasiholes, and states containing 
many quasiparticles and quasiholes at arbitrary filling factors.

\begin{acknowledgments}
Partial support by the National Science Foundation under grant
no. DMR-0240458 is gratefully acknowledged.
\end{acknowledgments}

\end{document}